\documentclass{article}

\usepackage{arxiv}

\usepackage[T1]{fontenc}    % use 8-bit T1 fonts
\usepackage{hyperref}       % hyperlinks
\usepackage{url}            % simple URL typesetting
\usepackage{booktabs}       % professional-quality tables
\usepackage{amsfonts}       % blackboard math symbols
\usepackage{nicefrac}       % compact symbols for 1/2, etc.
\usepackage{microtype}      % microtypography
\usepackage{lipsum}		% Can be removed after putting your text content
\usepackage{graphicx}
\usepackage{natbib}
\usepackage{doi}

\usepackage{makecell}
\usepackage[utf8x]{inputenc}
\usepackage{tfrupee}
\title{\textit{Sangyadhi}: An Analytical Framework for Diagnosing and Investigating Collective Care}
\renewcommand{\shorttitle}{\textit{Sangyadhi}}

\author{ \href{https://orcid.org/0000-0001-5787-4858}{\includegraphics[scale=0.06]{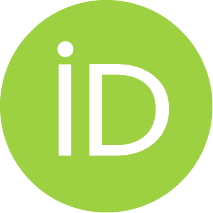}\hspace{1mm}Shyama Sastha~Krishnamoorthy Srinivasan}\\
	IIIT-Delhi\\
	New Delhi, India\\
	\texttt{shyamas@iiitd.ac.in} \\
    \And
    \href{https://orcid.org/0000-0002-0286-6997}{\includegraphics[scale=0.06]{orcid.pdf}\hspace{1mm}Mohan~Kumar}\thanks{Both authors are advisors of this work.}\\
	RIT, NY\\
	Rochester, New York, US\\
	\texttt{mjkvcs@rit.edu} \\
	\And
	\href{https://orcid.org/0000-0003-2152-1027}{\includegraphics[scale=0.06]{orcid.pdf}\hspace{1mm}Pushpendra~Singh}\footnotemark[1]\\
	IIIT-Delhi\\
	New Delhi, India\\
	\texttt{psingh@iiitd.ac.in} \\
}

\hypersetup{
pdftitle={Sangyadhi: An Analytical Framework for Diagnosing and Investigating Collective Care},
pdfsubject={cs.HC, cs.CY},
pdfauthor={Shyama Sastha~Krishnamoorthy Srinivasan, Mohan~Kumar, Pushpendra~Singh},
pdfkeywords={Collective Care, Health Informatics, Analytical Framework, Theory \& Methodology}
}

\begin{document}
\maketitle

\begin{abstract}
	While HCI has expanded from Personal to Family Informatics, researchers still lack a critical analytical vocabulary to diagnose what truly constitutes a collective care context. We introduce \textit{Sangyadhi}, an analytical sociotechnical framework that synthesizes Situated Action, Distributed Cognition, the Capabilities Approach, and Actor-Network Theory to investigate multi-actor care ecologies. The framework articulates four interdependent analytical lenses: contextual grounding, distributed responsibility, trust-mediated sharing, and collaborative actionability. To demonstrate its diagnostic utility, we apply \textit{Sangyadhi} to critically audit a decade of HCI digital health literature across self-tracking, family informatics, and clinical AI\textemdash surfacing structural omissions in how shifting authority, dynamic trust, and AI escalation are conceptualized. Ultimately, the goal of \textit{Sangyadhi} is to equip researchers with an analytical blueprint for diagnosing observed care contexts, structuring field inquiries, and evaluating future technologies.
\end{abstract}

% keywords can be removed

\section{Introduction}
\label{intro}

Most care technologies\textemdash from commercial digital health apps to personal informatics (PI) tools\textemdash assume a quietly powerful premise: the user is an independent actor working with an individual device that pushes data toward a personal dashboard. But because care is fundamentally relational, this premise is analytically impoverished; it produces brittle systems and distorts evaluation by measuring individual compliance rather than collective efficacy. In many real-world multigenerational households, neighborhood care networks, and community health programs, care is not an individual endeavor but an ongoing set of social arrangements. Monitoring, interpretation, and intervention are distributed across relatives, neighbors, and frontline health workers; rituals and negotiated norms govern timing and visibility; and authority is porous and shifting. Stage-based, individual-centered models of health informatics capture significant instrumental problems, but they do not provide researchers with the analytical tools needed to understand how the social, temporal, and moral economies of care are actually enacted. We refer to these interlocking realities as \textit{Collective Care}. Recent empirical work by Srinivasan et al.~\cite{srinivasan2025unpacking} demonstrates that in many sociocultural contexts, personal health informatics are inherently appropriated to support proactive collective care across families and communities. Yet, while consumer health technologies like Apple Health, Clue, and Flo have begun adding mechanisms for relational data sharing, the HCI community still lacks a critical analytical methodology to diagnose what fundamentally constitutes a collective care context, characterize its structural tensions, and guide empirical field inquiry into it. 

\begin{figure}[ht]
\centering
\includegraphics[width=.95\linewidth]{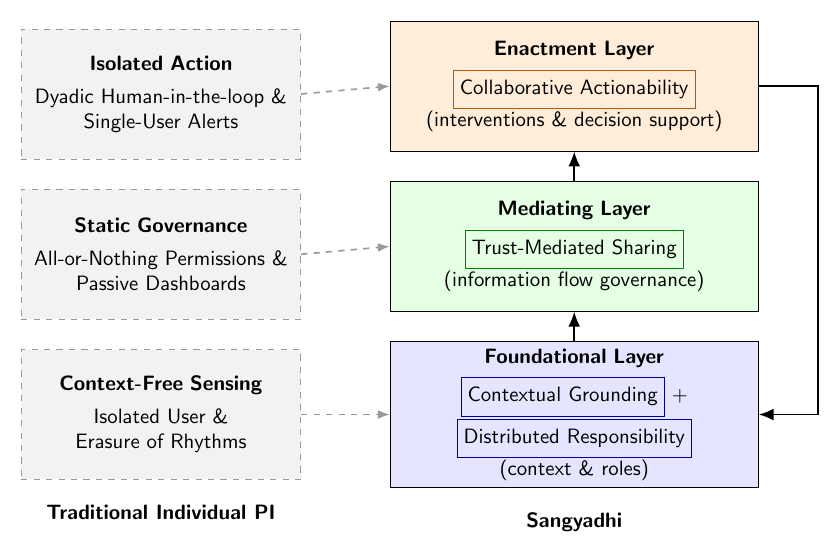}
\caption{\label{fig:arch} Conceptual architecture of the \textit{Sangyadhi} framework, contrasting traditional individual-centered paradigms (on the left) with the interdependent hierarchy of collective care ecologies (one the right).}
\end{figure}

This paper proposes \textit{Sangyadhi}\footnote{\textit{Sangyadhi} is a Sanskrit word derived from the combination of the words \textit{Sang} (meaning 'group' or 'collective') and \textit{dhi} (meaning 'understanding'). The naming implies a conscientious understanding of the collective nature of care.}, an analytical sociotechnical framework that reframes the primary unit of analysis from the isolated individual to the social arrangements that enable care. Rather than treating collective care as a special case or an ad hoc extension of individual health management, \textit{Sangyadhi} conceptualizes it as a situated, distributed, and relational activity shaped by people, technologies, norms, and infrastructures (Figure~\ref{fig:arch}). By doing so, the framework provides researchers with an analytical grammar for observing, diagnosing, and investigating these complex multi-actor environments.

While existing bodies of theory and empirical work richly describe isolated dimensions of collective activity\textemdash such as distributed cognition across artifacts~\cite{10.1145/353485.353487}, the situated nature of human agency~\cite{38407, emirbayer1998agency}, shifting power dynamics in families~\cite{10.1145/2998181.2998362, 10.1145/2998181.2998303, 10.1145/3274396}, and the relational flow of privacy~\cite{Nissenbaum_2004}\textemdash they largely operate as fragmented lenses. Stage-based models~\cite{10.1145/1753326.1753409} center the individual, infrastructuring scholarship~\cite{Star_Ruhleder_1996, Pipek_Wulf_2009} operates at a macro-scale, and human-AI interaction frameworks~\cite{10.1145/3290605.3300233} predominantly examine automation dyadically. The result is a persistent gap: researchers still lack an integrated conceptual synthesis capable of evaluating the shifting states of responsibility, dynamic information boundaries, and distributed action that define collective care.

What remains underdeveloped is a conceptual scaffolding capable of addressing four core questions central to multi-actor care environments. \textit{Sangyadhi} addresses this gap by synthesizing complementary theoretical perspectives into four interdependent research questions (RQs) that define our analytical constructs:

\begin{itemize}
\item \textbf{RQ1 (Contextual Grounding):} How are agency and care practices shaped by local domestic routines, cultural norms, familial relationships, and temporal conditions?
\item \textbf{RQ2 (Distributed Responsibility):} How are cognition, responsibility, and legitimate authority distributed, transferred, and renegotiated across actors and artifacts over time?
\item \textbf{RQ3 (Trust-mediated Sharing):} How do relational trust, vulnerability, and contextual boundaries shape actors' substantive capabilities to access, interpret, disclose, and control sensitive health information?
\item \textbf{RQ4 (Collaborative Actionability):} How do human and technological actors coordinate action, interpret system-generated insights, and negotiate judgment, deferral, escalation, and contestability across the care network?
\end{itemize}

Conceptually, \textit{Sangyadhi} synthesizes situated agency, distributed cognition, the capabilities approach, and actor-network theory into a coherent HCAI lens for investigating collective care. The constructs do not operate as isolated modules, but as interdependent analytical commitments: context shapes the conditions of agency; cognition and authority are distributed across care networks; information practices mediate actors' capabilities to participate; and action emerges through heterogeneous human-technological configurations. 

The objective of this work is threefold: (a) to synthesize complementary sociotechnical theories into four analytical constructs for collective care; (b) to demonstrate the diagnostic power of the framework by conducting a critical audit of the last decade of HCI digital health literature across individual self-tracking, family informatics, and clinical AI; and (c) to equip researchers with a concrete analytical blueprint and qualitative inquiry protocol for diagnosing observed field contexts, structuring empirical studies, and evaluating future collective care technologies.
\section{Why a Collective Care Framework is Needed?}
\label{literature}

Before presenting the proposed collective care framework, it is necessary to establish what existing approaches already explain about health technology use and where they remain analytically insufficient for investigating collective care contexts. This section, therefore, traces the evolution of several relevant bodies of work rather than treating them as static or uniformly inadequate. We first examine how Personal Informatics (PI) established an influential individual-centered model of health information work, and how subsequent research progressively extended it to include visualization, personalization, reflection, and behavioral support. We then consider research on family health monitoring and caregiving, which demonstrates that interpretation, coordination, and action frequently extend beyond the individual, while noting that these studies often adapt rather than fundamentally reframe the underlying unit of analysis. We next examine socio-cognitive and infrastructuring perspectives, which explain why care is distributed across people, artifacts, routines, and institutions. We then turn to privacy, trust, and value-sensitive approaches, followed by human-centered AI (HCAI) and automation research, to examine how information and algorithmic action are conceptualized when care involves multiple actors. Across these bodies of work, the recurring issue is not a lack of attention to sociality, collaboration, or ethics in isolation; rather, it is the absence of an integrated analytical account of how context, responsibility, information participation, and action interact in the collective enactment of care. Section~\ref{gap} synthesizes these gaps and derives \textit{Sangyadhi}'s theoretical commitments.

\subsection{Reorienting Personal Informatics: From Individual Self-tracking to Collective Care}
\label{pi}

Personal Informatics (PI) systems are commonly organized around a stage-based workflow of preparation $\rightarrow$ collection $\rightarrow$ integration $\rightarrow$ reflection $\rightarrow$ action~\cite{10.1145/1753326.1753409}. This model established a powerful vocabulary for understanding self-tracking by positioning the individual user as the primary agent who gathers, interprets, and acts on personal data. Subsequent research substantially expanded the capabilities of this individual-centered paradigm rather than abandoning it. Work on visualization and interpretation, for example, has improved how users make sense of increasingly complex health information. Liang et al.~\cite{10.1145/3544549.3585604} develop user-centered visualizations for interpreting multimodal outputs, while Lim et al.~\cite{10.1145/3706598.3714058} propose domain-aligned explanation techniques for clinical contexts. Other work focuses on behavior change and personalization. Hsieh et al.~\cite{10.1145/2559206.2560474} examine how persuasive technologies can adapt interventions to individual contexts, and related studies~\cite{10.1145/3411764.3445619, 10.1145/3706598.3713224, 10.1145/3706598.3714208, 10.1145/3706598.3714072, 10.1145/3706598.3713650} explore tailoring feedback to stages of change and motivational factors. Longitudinal analyses further identify challenges with sustained engagement; for instance, Kovács et al.~\cite{10.1145/3411764.3445695} report that intervention adherence often declines over time. Additional work examines heterogeneity in user characteristics~\cite{10.1145/3613904.3642352} and the use of multisensory feedback devices to influence behavior~\cite{10.1145/3613904.3642651}. Research in caregiver and dementia contexts has further demonstrated the importance of social support structures and equity considerations in tech design~\cite{10.1145/3411764.3445559, ma2024bridginggapadvancementstechnology}, alongside design probes and toolkits that highlight the emotional and temporal dimensions of informal caregiving~\cite{Bosch_Kanis_Dunn_Stewart_Kröse_2019, 10.1145/2851581.2892354}.

This progression shows that PI has become considerably more sophisticated in supporting individual interpretation, personalization, and action. At the same time, the dominant unit of analysis has generally remained the individual user. While consumer health platforms have increasingly added features for relational sharing, these extensions predominantly treat social interaction as a secondary form of sharing an individual's personal metrics. Empirical work on family health monitoring has begun to expose the limits of that assumption. Studies of family health monitoring systems demonstrate that data interpretation and decision-making are routinely distributed across household members rather than centered on a single user~\cite{10.1145/2998181.2998362}. Kaziunas et al.~\cite{10.1145/2998181.2998303} illustrate this dynamic in pediatric diabetes care, where different actors generate data, interpret it, and negotiate interventions. Similar patterns emerge across diverse care settings: monitoring tools can inadvertently shift coordination and emotional labor onto caregivers or relatives, sometimes increasing rather than relieving burden~\cite{10.1145/3274396, Abdelaziz_Garfield_Neves_Lloyd_Norton_van_Dael_Wheeler_McLeod_Franklin_2024, 10.1145/3715336.3735746}. The state-based medication routine framework~\cite{10.1145/3491102.3517519} further demonstrates that reminder-based systems falter when everyday social contingencies disrupt individually framed routines. Studies of sensemaking and reflection in PI likewise highlight breakdowns in contextualization when authority, experience, and responsibility are distributed among multiple actors~\cite{10.1145/3491102.3501991, 10.1145/3173574.3173775}.

Taken together, these studies do more than demonstrate that people often manage health collaboratively; they reveal a structural limitation in the individual-centered model itself. Family informatics and caregiver-oriented research frequently extend individual-centered tools to accommodate secondary participants, but they do not provide an analytical framework for settings in which the stages of sensing, interpretation, reflection, and action are decoupled among distinct actors. Thus, the analytical challenge is not merely that personal informatics tools lack social features. Rather, the field lacks a conceptual model that treats the collective arrangement\textemdash with its shifting division of labor, fluid authority, and situated visibility\textemdash as the primary unit of analysis.

The resulting gap for collective care inquiry is threefold: existing PI models under-specify \textit{who} participates in interpretation and action, \textit{how responsibility transitions} among participants, and \textit{under what circumstances} information becomes visible or actionable to particular actors. Existing family and caregiving studies provide strong empirical evidence for these friction points, but they primarily document coordination breakdowns within systems built for individuals. They do not provide an integrated conceptual vocabulary for diagnosing collective care as an interdependent sociotechnical practice in its own right.

\subsection{Socio-cognitive and Infrastructuring Foundations}
\label{socio-cognitive}

The empirical literature on family health and caregiving points to a broader theoretical question: how should researchers conceptualize and evaluate activities distributed across people, artifacts, routines, and environments? Several theoretical traditions provide conceptual tools for understanding this problem. Distributed cognition research emphasizes that cognitive processes are not confined to individuals but are distributed across people, artifacts, and environments. Hollan, Hutchins, and Kirsh~\cite{10.1145/353485.353487} describe how interaction with external artifacts and other actors forms a cognitive system that extends beyond individual minds. Activity theory similarly conceptualizes human activity as a structured system of actors, tools, rules, and shared objectives; Engeström~\cite{engeström2015learning} uses this framework to explain how collective practices evolve and transform over time. Social practice theory complements these perspectives by emphasizing how routines stabilize through the interaction of materials, competences, and meanings~\cite{shovesocial}.

Infrastructuring research extends these ideas by examining how people assemble and adapt sociotechnical arrangements in everyday contexts. Star and Ruhleder~\cite{Star_Ruhleder_1996} analyze how infrastructures emerge through ongoing use rather than through singular design decisions, while Pipek and Wulf~\cite{Pipek_Wulf_2009} and Dalsgaard and Dindler~\cite{10.1145/2556288.2557342} conceptualize infrastructuring as a continuous design process bridging theory and practice. Empirical studies demonstrate how communities repurpose existing technologies to support coordination and care. For example, Wilcox et al.~\cite{10.1145/3544548.3581040} document how marginalized communities adapt communication tools and informal artifacts to support wellbeing. Monteiro et al.~\cite{10.1007/s10606-012-9167-1} describe how artifacts gradually become infrastructural as they are embedded in everyday practices. These reflections on infrastructure clarify how conventions, access arrangements, and evolving practices shape long-running sociotechnical systems.

These perspectives are especially important for collective care because they challenge the assumption that care work can be understood as the execution of tasks by a single autonomous user. They establish that knowing, coordinating, and acting depend on relations among people and artifacts, as well as on routines that emerge over time. They also help explain why informal communication channels, household practices, community resources, and repurposed technologies become essential parts of care infrastructures.

However, distributed cognition and activity theory do not by themselves distinguish between the \textit{distribution of cognitive work} and the \textit{distribution of legitimate responsibility}. A system or household arrangement may help several people contribute information without clarifying who is currently responsible for interpreting it, who holds the authority to decide, or who has accepted accountability for acting. Similarly, infrastructuring accounts explain how arrangements become embedded and adapted in practice, but provide limited analytical scaffolding for evaluating these changing relationships within interactive digital systems. The collective care gap is therefore not a lack of theories of distributed activity; it is the lack of an analytical account of how shifting responsibility, authority, and accountability can be diagnosed and made negotiable within those distributed arrangements.

This creates an analytical-diagnostic gap. Existing theories richly explain how distributed work unfolds descriptively, but they offer limited diagnostic guidance for investigating role-aware handoffs, dynamic notification routing, responsibility transfer, or mutual accountability in digital environments. For collective care research, our frameworks must not only recognize that activity is distributed; they must also provide the analytical criteria to evaluate how responsibility is allocated, transferred, contested, and re-established as the care situation evolves. This motivates the need for a construct that extends the insight of distributed cognition from the distribution of knowing and doing toward the distribution and renegotiation of responsibility.

\subsection{Privacy, Trust, and Value-sensitive Approaches to Care}
\label{privacy}

Privacy and values scholarship provides another important foundation for studying technologies that mediate sensitive health information. Contextual Integrity conceptualizes privacy as the appropriate flow of information within specific social contexts~\cite{Nissenbaum_2004}. Value-Sensitive Design (VSD) extends this perspective by emphasizing the importance of explicitly incorporating stakeholder values into technological design processes~\cite{HCI-015}. Researchers have applied these frameworks to investigate governance mechanisms for health data systems. Studies on role-based access models examine how different caregiving roles should correspond to distinct privileges and responsibilities within health platforms~\cite{10.1145/3613904.3642728}. Other work investigates consent interfaces and policy designs that aim to give patients and families clearer control over data sharing~\cite{10.1145/3613904.3642815, 10.1145/3613904.3642851, 10.1145/3706598.3714089, 10.1145/3706598.3714019}. Audit trails and logging mechanisms have also been proposed as ways to increase transparency and accountability in health data exchanges~\cite{10.1145/3706598.3714019}.

These approaches are particularly relevant to collective care because information does not move solely among abstract ``users.'' It moves through social relationships in which the legitimacy of access depends on role, purpose, trust, timing, and responsibility. Despite the advances described above, traditional access control frameworks~\cite{485845, 10.1145/1943513.1943539} reveal limitations when access is evaluated primarily as a static configuration. Research shows that caregiving relationships inherently require dynamic access arrangements. For example, caregivers often require temporary emergency access to information that must later be revoked~\cite{10.1145/2441776.2441838}. Studies of family health systems similarly document situations where relatives or caregivers temporarily assume interpretive authority during acute health events~\cite{10.1145/2998181.2998303}. Work examining community and institutional health interventions further highlights the need for bounded, auditable access for intermediaries such as community health workers~\cite{10.1145/3274344, 10.1145/3274345} and medical teams~\cite{10.1145/2531602.2531643, 10.1145/2675133.2675279, 10.1145/2998181.2998328}. Deployments in low-resource contexts also demonstrate how social norms and practical constraints shape how and when health information is shared~\cite{10.1145/3313831.3376170}.

Broader theoretical work reinforces these observations. Ostrom's theory of collective governance~\cite{ostrom1993design} argues that shared resource systems require rules and monitoring mechanisms that make cooperation observable and enforceable. Feminist HCI critiques further highlight how technological systems can encode power relations and reproduce inequalities if social dynamics are ignored~\cite{10.1145/3706598.3713950}. Standpoint theory similarly emphasizes the importance of centering marginalized caregivers whose experiences reveal overlooked analytical needs~\cite{Hartsock1983}.

Empirical studies of digital health adoption further demonstrate how structural barriers and gendered roles shape technology use~\cite{chronaki2017search}. Related studies emphasize the importance of culturally sensitive perspectives that support caregivers within their social contexts~\cite{O_Connor2025, xu2025impact, october2023_gender_digital_health, s2025whotimeunderstandingreceptivity, choi2025bridging}. The concept of ``caring-through-data'' reframes health information practices as relational activities embedded within social ties and obligations~\cite{10.1145/2998181.2998303}. Methodologically, value-oriented design~\cite{PEREIRA201566}, alongside participatory and action-oriented research traditions, offers ways to engage these complexities during inquiry. Participatory Action Research (PAR) emphasizes collaborative inquiry and iterative reflection~\cite{Baum854}. Ethnographic Action Research (EAR) extends these principles through long-term engagement with communities and iterative refinement~\cite{10.1108/IJMPB-09-2021-0232, BetterEvaluation_EARToolbox_2017, black2021ethnographic, morgan2016ethnographic}. Recent HCI work similarly demonstrates how participatory design and reflexive ethnography can ground care technologies in participants' lived routines and values~\cite{10.1145/3706598.3713436, 10.1145/3613904.3642727, 10.1145/3706598.3713755}.

Together, these perspectives establish that privacy, trust, and participation in caregiving are negotiated, relational, and evolving. Yet this literature leaves an important gap for collective care inquiry. Existing privacy- and value-sensitive approaches explain why information flows must respect context, values, and relationships, while empirical and participatory studies show how those conditions vary in practice. What remains underdeveloped is an analytical understanding of how information visibility dynamically fluctuates as caregiving roles, purposes, and responsibilities change. In collective care, the appropriate recipient of information cannot be evaluated solely through a fixed role or a one-time consent decision. A caregiver may need temporary access during an acute episode; a clinician may require different information from a family member; and an actor's legitimate need to know depends directly on whether they are currently responsible for action. The analytical challenge is therefore not simply to evaluate ``better privacy,'' but to understand the systematic integration of information visibility with the shifting structure of collective responsibility.

\subsection{AI, Automation, and Human-in-the-loop in Care}
\label{ai-care}

HCAI research emphasizes the importance of transparency, explainability, and the maintenance of meaningful human oversight in automated systems~\cite{10.1145/3290605.3300233}. Shneiderman argues for interaction paradigms that keep humans in control of automated decision-making~\cite{shneiderman2022human}, and subsequent work outlines operational principles for applying these ideas within healthcare contexts~\cite{10.1145/3706598.3713184}.

Applied research has significantly advanced predictive and recommendation capabilities in health technologies. Studies have developed models for clinical risk prediction~\cite{islam2023machine}, anomaly detection in physiological sensor data~\cite{sunny2022anomaly}, and personalized recommendation systems for health management~\cite{cai2024machine, 10.1145/3397318, 10.1145/3544548.3581251, 10.1145/3706598.3713852}. These approaches improve the technical capacity of digital health systems to identify risks, detect anomalies, generate tailored recommendations, and calibrate XAI to user expertise~\cite{10.1145/3411764.3445088}. Recent advancements also explore the impact of outcome-feedback on user trust~\cite{10.1145/3613904.3642780}, prototypes for personalized LLM explanations~\cite{10.1145/3706599.3719982}, and the LLM-based demystification of complex clinical reports~\cite{10.1145/3706599.3720208}.

However, HCI research also documents persistent challenges when automation is integrated into real-world workflows. Studies report automation bias and over-reliance on algorithmic outputs~\cite{10.1109/ICHI.2015.26}. Other work highlights difficulties incorporating algorithmic advice into everyday decision-making practices~\cite{10.1145/3491102.3502104, 10.1145/3610218}. Recent analyses examine how automation can shift coordination work or introduce new forms of invisible labor~\cite{ulloa2022invisible}, and how poorly contextualized automation may generate unintended harms~\cite{10.1145/3643834.3660707, kucking2024automation}.

In collective care settings, these challenges are compounded because algorithmic outputs circulate within multi-actor ecologies where participants hold diverse roles, clinical literacies, and decision-making authority~\cite{10.1145/3706598.3713503}. Field studies of smart-home and health deployments demonstrate that automation that lacks awareness of relational dynamics can disrupt household equilibrium and impose acute social burdens~\cite{10.1145/3544548.3581546, 10.1145/3544548.3581553}. These tensions expose a structural limitation in conventional HCAI framing: current frameworks predominantly model interaction as a \textit{dyadic loop} between an automated agent and an isolated decision-maker. In collective care, the analytical question is not simply whether ``a human'' remains in control, but how automated outputs are interpreted, negotiated, and enacted within an interdependent social network in which authority and availability are continuously shifting.

Data standards and integration frameworks also expose the distinction between technical availability and collective action. Efforts to standardize health data exchange, such as FHIR-based interoperability approaches~\cite{10.1145/3402524}, and conceptual models such as the Human Data Model~\cite{10.1145/3706598.3714069}, primarily address data portability rather than the coordination and social governance of caregiving activities. As a result, questions about how automated insights should be routed, deferred, escalated, or contested within caregiving networks remain analytically under-theorized.

The resulting collective care gap is therefore an analytical gap in \textit{actionability}. Existing HCAI and automation research provides substantial guidance on transparency, explanation, and human oversight, while empirical HCI work demonstrates the risks of automation that ignores social relationships. Yet these approaches do not fully specify how AI outputs become actionable across a distributed care network in which responsibility can be shared, transferred, or contested. What is needed is an analytical account of actionability that investigates not only the generation of a recommendation, but also its routing, interpretation, deferral, escalation, and enactment across the human and technological actors that collectively constitute care.

\subsection{Theoretical Gap and Construct Derivation}
\label{gap}

The preceding literature does not suggest that existing approaches are generally inadequate. Rather, each addresses an important dimension of care while leaving a particular aspect of collective care underspecified. PI research establishes increasingly sophisticated mechanisms for sensing, reflection, personalization, and action, but its dominant unit of analysis remains the individual user; family informatics and caregiving research demonstrates that these stages are often distributed across actors, while also revealing burden, coordination problems, and contextual friction that individual-centered models do not adequately capture. Distributed cognition, activity theory, social practice theory, and infrastructuring explain how knowing, doing, and coordination are distributed across people, artifacts, routines, and environments, but do not provide an analytical framework for evaluating shifts in responsibility and authority. Privacy, contextual integrity, value-sensitive design, and related empirical work establish that information practices are relational, contextual, and value-laden, yet they do not fully connect information visibility to changing participation and responsibility within a care network. Advances in HCAI and automation research explainability, oversight, and responsible interaction, but collective care settings require an additional analytical account of how AI-generated outputs are routed, interpreted, deferred, contested, and enacted across multiple actors.

These observations reveal four interdependent questions that existing bodies of work tend to address separately:

\begin{enumerate}
\item \textit{Where is the agency situated?} Care actions depend on social, material, temporal, and relational circumstances rather than occurring independently of context.
\item \textit{How are responsibility and authority distributed?} Collective care distributes sensing, interpretation, decision-making, and action across people and artifacts, while responsibility and authority shift over time.
\item \textit{What enables actors to participate in information practices?} Appropriate participation depends on context, relationships, role, purpose, and the conditions under which information becomes visible and actionable.
\item \textit{How does information become coordinated action?} Care-related recommendations and decisions are enacted through heterogeneous networks of patients, caregivers, clinicians, technologies, representations, and institutional arrangements.
\end{enumerate}

These questions point to an analytical gap that is more specific than the absence of any one theory of collective activity. What is missing is an integrated sociotechnical vocabulary that connects \textit{context}, \textit{responsibility}, \textit{information participation}, and \textit{action} as mutually dependent dimensions of collective care. The empirical literature demonstrates that these dimensions repeatedly become entangled in practice: changes in context can alter who is responsible; changes in responsibility can alter who needs access to information; changes in information visibility can enable or constrain participation; and the resulting configuration of actors and technologies determines whether an AI-supported action can be appropriately enacted.

Recent fieldwork by Srinivasan et al.~\cite{srinivasan2025unpacking} crucially articulated the multi-level socio-ecological landscape of collective care, illustrating tensions spanning micro-, meso-, exo-, and macro-systems. However, while ecological models map the structural spheres within which care occurs, researchers still lack a cross-cutting analytical grammar to diagnose how operational dynamics\textemdash specifically contextual grounding, shifting responsibility, dynamic visibility, and multi-actor actionability\textemdash transact across these levels. To address this gap, \textit{Sangyadhi} synthesizes four complementary theoretical perspectives: situated action and agency; distributed cognition and activity theory; the capabilities approach and contextual information-flow theory; and actor-network theory, alongside HCAI. Following a review $\rightarrow$ analyze $\rightarrow$ critique approach to theory-driven HCI synthesis~\cite{10.1145/3613904.3642624}, we bring these traditions together to overcome analytical fragmentation. The resulting four constructs are theoretically synthesized to provide an integrated diagnostic framework for characterizing, investigating, and evaluating collective care ecologies in HCI.
\section{The Sangyadhi Framework}
\label{the-framework}

Building on the theoretical synthesis developed in Section~\ref{literature}, we introduce \textit{Sangyadhi} as an analytical sociotechnical framework for diagnosing, investigating, and conceptualizing collective care. The framework translates the four interdependent gaps identified across prior literature into an integrated diagnostic grammar. Table~\ref{tab:constructs} summarizes the theoretical grounding of these constructs, along with their corresponding diagnostic criteria and sociotechnical design levers. We define each construct, articulate its analytical role in characterizing collective care contexts, and show how its commitments inform empirical inquiry and system design.

\begin{table}[ht!]
\centering
\small
\begin{tabular}{@{}p{0.14\textwidth} p{0.25\textwidth} p{0.51\textwidth}@{}}
\toprule
\textbf{Construct} & \textbf{Theoretical grounding} & \textbf{Diagnostic Criteria \& Sociotechnical Levers} \\
\midrule
Contextual Grounding &
\textit{Situated action and agency}~\cite{38407, emirbayer1998agency} &
\begin{tabular}[t]{@{}p{\linewidth}@{}}
\textbf{Temporal and routine alignment:} Investigating how care routines intersect with domestic, cultural, and environmental rhythms; supporting context-aware timing rather than rigid alerts. \\
\textbf{Situated framing:} Examining how health events are socially interpreted; framing notifications and interactions to reflect local interpersonal dynamics. \\
\textbf{Contestable norms:} Evaluating whether encoded rules remain transparent, adaptable, and renegotiable by the care network rather than imposing rigid external standards.
\end{tabular} \\
\midrule

Distributed Responsibility &
\textit{Distributed cognition} and activity theory~\cite{10.1145/353485.353487, engeström2015learning} &
\begin{tabular}[t]{@{}p{\linewidth}@{}}
\textbf{Legibility of responsibility:} Tracing who holds current interpretive and operational accountability; distinguishing technical access from legitimate social authority. \\
\textbf{Handoff signaling:} Identifying how responsibility shifts, is delegated, or is contested across actors during acute episodes or routine care transitions. \\
\textbf{Differentiated coordination:} Evaluating whether multi-actor representations preserve distinct participant roles and identities without flattening them into a single user.
\end{tabular} \\
\midrule

Trust-mediated Sharing &
\textit{Capabilities approach} and contextual information-flow theory~\cite{sen1999development, Nussbaum_2011, Nissenbaum_2004} &
\begin{tabular}[t]{@{}p{\linewidth}@{}}
\textbf{Time-bounded visibility:} Examining how access boundaries fluctuate; evaluating purpose-bound and ephemeral data sharing with explicit boundaries. \\
\textbf{Progressive disclosure:} Assessing whether data granularity aligns with an actor's legitimate role, technical literacy, and relational proximity. \\
\textbf{Provenance transparency:} Tracing whether participants can inspect who generated, accessed, or transformed health information, supporting mutual trust and recourse.
\end{tabular} \\
\midrule

Collaborative Actionability &
\textit{Actor-network theory} and HCAI~\cite{Callon_1984, Latour_2005, 10.1145/3290605.3300233} &
\begin{tabular}[t]{@{}p{\linewidth}@{}}
\textbf{Role-aware routing:} Diagnosing whether system outputs reach the specific human actor positioned to act, rather than an undifferentiated dashboard. \\
\textbf{Deferral and escalation:} Identifying social and institutional pathways for holding, redirecting, or escalating alerts when primary actors are unavailable. \\
\textbf{Network-aware explanations:} Scrutinizing how explanations articulate the dependencies among distributed actors, prior interventions, and algorithmic inferences.
\end{tabular} \\
\bottomrule
\end{tabular}
\caption{\label{tab:constructs} The Sangyadhi framework: theoretical foundations, constructs, and diagnostic criteria paired with illustrative sociotechnical levers.}

\end{table}

\subsection{The Core Constructs}
\label{core}

\textit{Sangyadhi} articulates four interlocking constructs that translate its theoretical foundations into diagnostic and analytical commitments. Rather than operating as isolated functional modules, these four constructs function as an interdependent diagnostic grammar. A shift in one dimension dynamically reconfigures the others: changes in domestic routines alter who is positioned to assume responsibility; shifts in responsibility dictate who legitimately requires information visibility; and the resulting distribution of knowledge determines how coordinated action can unfold. Here, we delve deeply into each construct to address the research questions (RQs) posed in Section~\ref{intro}.

\subsubsection{Contextual Grounding}

Addressing \textbf{RQ1}, situated action and agency emphasize that human activity cannot be understood as the rote execution of pre-formulated plans, but is constituted through ongoing engagement with particular social, material, and temporal circumstances~\cite{38407, emirbayer1998agency}. In collective care, the legitimacy and efficacy of a health practice depend heavily on domestic rhythms, interpersonal relationships, resource constraints, and cultural expectations. An intervention or recommendation that is clinically sound in isolation may fail or cause harm if it ignores the situated conditions under which it must be enacted.

\textit{Contextual Grounding} establishes an analytical lens for investigating how care practices are shaped by and embedded within local environments. In field inquiry, this construct guides researchers to probe temporal structures (e.g., quiet windows, caregiving shifts, religious rituals), household hierarchies, and implicit communicative etiquette. For sociotechnical systems, Contextual Grounding requires that technologies make these conditions visible and negotiable rather than imposing static assumptions. By treating context as an active, situated constituent of care rather than an external variable, this lens helps researchers evaluate whether technologies align with, or disrupt, the lived realities of caregiving networks.

\subsubsection{Distributed Responsibility}

Addressing \textbf{RQ2}, distributed cognition conceptualizes cognitive work as extending across people, artifacts, and environments~\cite{10.1145/353485.353487}, while activity theory articulates how collective practice is structured by rules, community norms, and division of labor~\cite{engeström2015learning}. In collective care settings, sensing, remembering, interpreting, deciding, and acting are rarely concentrated within an individual; they are distributed across an ensemble of informal and formal actors.

\textit{Distributed Responsibility} extends these cognitive perspectives to examine the social organization of accountability and authority. An analytical limitation of existing health systems is that they frequently conflate technical capability (e.g., who has access to an app) with legitimate authority (e.g., who is culturally or clinically sanctioned to make a care decision). This construct equips researchers to trace how responsibility is allocated, how handoffs occur when primary caregivers are unavailable, and where invisible coordination labor creates friction. Diagnostically, it asks: \textit{Who is accountable for interpreting an emerging trend? Who holds the authority to intervene? And how does the system make transitions of responsibility visible, contestable, and shared across the network?}

\subsubsection{Trust-mediated Sharing}

Addressing \textbf{RQ3}, the capabilities approach distinguishes between formal access and the substantive freedom individuals have to participate meaningfully in decisions affecting their well-being~\cite{sen1999development, Nussbaum_2011}. Contextual Integrity complements this by showing that privacy is not a blanket right to secrecy, but the appropriate flow of information relative to context-specific norms and relationships~\cite{Nissenbaum_2004}. In collective care, health information is deeply relational; its disclosure can cultivate solidarity or provoke surveillance, shame, and interpersonal tension.

\textit{Trust-mediated Sharing} operationalizes these insights into an analytical lens for evaluating how information flows govern participation. Rather than analyzing privacy as a static binary toggle (share vs. do not share), this construct evaluates sharing as a dynamic, purpose-bound, and relational negotiation. It directs researchers to investigate how boundaries of visibility adapt during acute health episodes versus during routine monitoring, how data granularity is tailored to participants' roles, and how provenance transparency enables actors to understand who accessed or altered records. Trust is thus analyzed not as an innate user attitude, but as an ongoing condition maintained through transparent, contestable, and bounded information practices.

\subsubsection{Collaborative Actionability}

Addressing \textbf{RQ4}, actor-network theory emphasizes that agency and action emerge through heterogeneous networks in which human and technological artifacts mediate and translate one another's activities~\cite{Callon_1984, Latour_2005}. HCAI scholarship reinforces the need for automated systems to preserve meaningful human agency, judgment, and oversight~\cite{10.1145/3290605.3300233}. In multi-actor care, an algorithmic inference or clinical recommendation is never an isolated endpoint; it becomes actionable only through the distributed network of people, cultural norms, communication channels, and institutional interfaces that receive and enact it.

\textit{Collaborative Actionability} provides an analytical framework for investigating how data and automated insights are translated into coordinated intervention. This construct examines the pathways through which recommendations are routed, interpreted, contested, deferred, or escalated. It challenges the conventional dyadic ``human-in-the-loop'' model by investigating \textit{which} human within a care network is positioned to act, under what relational constraints, and with what explanatory support. Diagnostically, this lens exposes failures where systems generate alerts without clear escalation pathways, burden peripheral actors with clinical alarms, or fail to provide mechanisms for humans to contest or defer automated advice.

\subsection{Architectural Overview}
\label{overview}

\textit{Sangyadhi} organizes these constructs into a layered sociotechnical model (see Figure~\ref{fig:arch}) that reflects the structural dependencies of collective care ecologies. 

\begin{itemize}
\item The \textbf{Foundational Layer} comprises \textit{Contextual Grounding} and \textit{Distributed Responsibility}. It establishes the baseline social conditions of care: the domestic routines, cultural practices, temporal rhythms, and division of authority that structure the care network. Without an explicit understanding of these foundational arrangements, any technological intervention risks operating on distorted assumptions about agency and duty.
\item The \textbf{Mediating Layer} comprises \textit{Trust-mediated Sharing}. It governs how information flows across the relationships established in the foundational layer. This layer regulates data visibility, progressive disclosure, purpose-bounded consent, and provenance, ensuring that access to sensitive health information is calibrated to an actor's legitimate role and current state of responsibility.
\item The \textbf{Enactment Layer} comprises \textit{Collaborative Actionability}. It represents the operational space where information, algorithmic insights, and human judgments are transformed into coordinated health interventions. This layer encompasses role-aware routing, contextual deferral, escalation paths, and contestability mechanisms.
\end{itemize}

These layers are analytically distinct but functionally recursive: foundational domestic rhythms dictate who assumes responsibility; responsibility structures determine the visibility of legitimate information in the mediating layer; and the resulting visibility enables coordinated interventions at the enactment layer, which in turn feed back to alter household routines and responsibility states. Rather than describing a software pipeline, this layered architecture provides researchers with an analytical map to locate the origins of sociotechnical breakdowns within a care ecology.

\subsection{Diagnostic Dimensions and Sociotechnical Implications}
\label{governance}

To operationalize the four constructs for field research and technology evaluation, \textit{Sangyadhi} identifies core diagnostic dimensions. These dimensions serve as analytical probes for qualitative inquiry and heuristic evaluation, while providing high-level direction for sociotechnical design.

\subsubsection{Contextual Grounding}
Researchers should investigate how temporal, relational, and cultural conditions shape health management. Diagnostically, this entails examining whether care routines follow predictable schedules or emergent events, identifying domestic rituals and quiet windows, and surfacing implicit cultural expectations. For system design, this implies moving away from rigid, clock-based notifications toward context-aware, routine-aligned interactions that allow households to configure, adapt, and contest encoded rules.

\subsubsection{Distributed Responsibility}
Inquiry must differentiate between task performance, interpretive labor, and decision-making authority. Diagnostically, researchers should map who records data, who interprets trends, who is authorized to initiate clinical escalations, and how handoffs occur during emergencies or caregiver fatigue. For technology design, this highlights the need for explicit representations of responsibility states, lightweight handoff signaling, and coordination views that support multiple participants without collapsing distinct roles into an undifferentiated user account.

\subsubsection{Trust-mediated Sharing}
Researchers should trace the boundaries, vulnerabilities, and relational dynamics governing data exchange. Diagnostically, this involves evaluating whether data sharing is treated as an all-or-nothing toggle, identifying boundaries of modesty or stigma, and examining how access needs shift during acute crises. For system design, this points toward dynamic, purpose-bound access models, progressive disclosure calibrated to recipient literacy and role, and transparent provenance logs that clarify who accessed or modified data.

\subsubsection{Collaborative Actionability}
Inquiry should focus on how insights and recommendations circulate through the multi-actor network. Diagnostically, researchers should analyze how alerts are routed, whether recommendations match recipient authority, and what opportunities exist for actors to contest, defer, or override automated suggestions. For technology design, this emphasizes multi-actor routing, contextual deferral mechanisms that respect domestic quiet hours, synchronized explanations that support collaborative discussion, and clear escalation protocols.

Together, these dimensions provide researchers with a principled vocabulary for interrogating collective care. By analyzing how context, responsibility, information flows, and actionability interact, \textit{Sangyadhi} enables the HCI community to move beyond surface-level observations of social sharing to diagnose the health of collective care arrangements.
\section{Applying Sangyadhi: Diagnostic Audits and Field Inquiry Protocols}
\label{application}

To demonstrate the analytical and generative utility of \textit{Sangyadhi}, this section illustrates how the framework functions as both a diagnostic lens for evaluating digital health systems and a methodological grammar for structuring empirical field inquiry. While Section~\ref{literature} surveyed existing literature to identify conceptual gaps, here we operationalize \textit{Sangyadhi} in three complementary stages: first, we conduct a structured diagnostic audit across three dominant research paradigms in HCI digital health and their corresponding platform archetypes; second, we present an illustrative diagnostic walkthrough of an urban postpartum caregiving setting to demonstrate how researchers can map, characterize, and evaluate multi-actor dynamics; and third, we synthesize these analytical procedures into a practical field inquiry protocol and qualitative diagnostic matrix for investigating collective care in the wild.

\subsection{Diagnostic Paradigm Audit: Interrogating Digital Health Paradigms and Ecosystems}
\label{critique}

To evaluate how digital health systems conceptualize\textemdash or critically overlook\textemdash the relational dynamics of care, we apply \textit{Sangyadhi}'s four constructs as a diagnostic audit across three major paradigms in the HCI literature. For each paradigm, we examine how theoretical premises shape concrete technological implementations by interrogating our core research questions (RQ1-RQ4). Specifically, we ask: how are domestic temporalities acknowledged, how is legitimate decision-making authority allocated, how are relational boundaries governed, and how are automated clinical insights routed across the care network?

\subsubsection{Paradigm 1: Individual-Centered Personal Informatics and Self-Tracking}
The foundational paradigm of Personal Informatics (PI) is organized around stage-based models of individual self-monitoring, reflection, and behavior change~\cite{10.1145/1753326.1753409}. Subsequent research has considerably refined this individual-centered model, investigating complex metric ecologies~\cite{10.1145/3706598.3713650}, deeper modalities of reflection~\cite{10.1145/3491102.3501991}, and tailored behavior change interventions~\cite{10.1145/3706598.3714072, 10.1145/3411764.3445695}. In commercial health technologies, this paradigm is archetyped by platforms like Apple HealthKit, which aggregate continuous biometric streams into a centralized repository governed by per-data-type permissions and provider integrations~\cite{Apple_HealthKit, Apple_HealthKit_Authorization, Apple_HealthKit_Records}.

\textit{Diagnostic Critique through Sangyadhi:} Audited through our framework, the individual-centered paradigm reveals critical structural omissions. In terms of \textit{Contextual Grounding}, self-tracking models treat sensor streams as context-free, objective physiological signals, erasing household rhythms, domestic rituals, and cultural timing that determine when data collection or health interventions are socially appropriate~\cite{10.1145/3715336.3735746}. Most fundamentally, the paradigm conflates the stages of informatics: it presumes that the individual who \textit{generates} the data (collection) is identical to the actor who \textit{interprets} its meaning (reflection) and executes the intervention (action). In collective care, however, these stages fracture across an ensemble: one member generates data, another interprets its significance, a third holds cultural or financial authority to decide, and a fourth is tasked with executing care~\cite{10.1145/2998181.2998303}. In terms of \textit{Distributed Responsibility}, platforms like Apple HealthKit support sharing only as a passive, read-only mirror~\cite{Apple_HealthKit_Authorization}. Granting secondary visibility does not establish who is accountable for addressing an abnormal trend, nor does it provide mechanisms for temporary handoffs of responsibility when a primary caregiver is indisposed. Consequently, individual-centered systems render invisible the immense coordination labor required to translate personal data into collective action.

\subsubsection{Paradigm 2: Relational and Interpersonal Informatics}
Responding to the solitary focus of self-tracking, HCI expanded toward Family Informatics and socially enabled health systems~\cite{10.1145/2998181.2998362, 10.1145/3274396, 10.1145/3674504}. This body of literature highlights how health tracking intersects with family communication~\cite{10.1145/2998181.2998362}, storytelling and social modeling~\cite{10.1145/3674504, 10.5555/AAI10223295}, informal caregiving roles~\cite{10.1145/3613904.3642728, 10.1145/2441776.2441789}, and the emotional vulnerabilities of ``caring-through-data''~\cite{10.1145/2998181.2998303}. In industry, this shift is embodied by relational features in specialized health applications, such as Flo for Partners~\cite{Flo_Partners, Flo_Partners_Privacy}, Clue Connect~\cite{Clue_Connect}, and Google Health Connect's cross-application interoperability platform~\cite{Google_HealthConnect, Google_HealthConnect_DataTypes, Google_HealthConnect_Architecture, Google_HealthConnect_Write}.

\textit{Diagnostic Critique through Sangyadhi:} While relational informatics successfully acknowledges social presence, \textit{Sangyadhi} reveals that it still lacks an operational vocabulary for dynamic governance, shifting authority, and multi-party accountability. Commercial ecosystems predominantly reduce relational care to \textit{static pairwise sharing} (e.g., a pregnant person to a partner or an aging parent to an adult child)~\cite{Flo_Partners, Clue_Connect}. Through the lens of \textit{Trust-mediated Sharing}, these implementations treat data access as an uncalibrated, permanently open pipeline, failing to accommodate the time-bounded, purpose-specific, and ephemeral disclosure arrangements required in real-world caregiving~\cite{10.1145/2441776.2441838, 10.1145/3706598.3714089, 10.1145/3613904.3642851}. In terms of \textit{Distributed Responsibility}, platforms like Google Health Connect achieve technical data exchange across disparate applications~\cite{Google_HealthConnect_Architecture}, but remain entirely role-agnostic: they provide no mechanisms to represent who is on duty, who has accepted responsibility for a task, or how handoffs are negotiated. Furthermore, relational systems frequently assume an idealized, harmonious domestic unit, overlooking how unmediated data sharing can reinforce patriarchal dynamics, trigger surveillance anxiety, or exacerbate gendered caregiving burdens~\cite{10.1145/3544548.3581546, 10.1145/3706598.3713950, 10.1145/3706598.3714019}.

\subsubsection{Paradigm 3: Clinical Decision Support and Human-Centered AI (HCAI)}
The rapid deployment of machine learning and generative AI in healthcare has driven substantial research in Clinical Decision Support Systems (CDSS) and HCAI frameworks~\cite{10.1145/3290605.3300233, shneiderman2022human}. This work focuses on algorithmic risk prediction~\cite{islam2023machine, cai2024machine, sunny2022anomaly}, clinician trust and advice-taking~\cite{10.1145/3544548.3581251, 10.1145/3491102.3502104, 10.1109/ICHI.2015.26}, counterfactual and abductive explanations~\cite{10.1145/3610218, 10.1145/3706598.3714058}, and calibrating XAI to user expertise~\cite{10.1145/3411764.3445088}. Concurrently, field deployments in global health emphasize that introducing automation into clinical infrastructures often produces invisible labor and institutional friction~\cite{ulloa2022invisible, 10.1145/3643834.3660707, 10.1145/3544548.3581553, 10.1145/3274344}.

\textit{Diagnostic Critique through Sangyadhi:} Through the lens of \textit{Collaborative Actionability}, HCAI research suffers from a persistent \textit{dyadic blind spot}: it overwhelmingly conceptualizes decision-making as a closed loop between an individual human user and an algorithmic agent~\cite{shneiderman2022human, 10.1145/3290605.3300233}. In collective care environments, algorithmic outputs do not terminate at the individual; they must circulate through complex social networks comprising patients, family members, and frontline intermediaries (such as community health workers or visiting nurses)~\cite{10.1145/3274344, 10.1145/3274345}. Current HCAI models lack analytical and operational mechanisms for \textit{role-aware routing} (directing recommendations to the actor positioned to act), \textit{contextual deferral} (pausing non-urgent alerts during domestic rest windows), and \textit{network-aware explanations} (clarifying how an algorithmic recommendation relates to the division of household labor). Without these mechanisms, automated interventions generate automation bias, alarm fatigue, or familial conflict by broadcasting urgent clinical warnings to individuals who lack the clinical literacy, economic autonomy, or domestic authority to resolve them~\cite{10.1109/ICHI.2015.26, kucking2024automation, s2025whotimeunderstandingreceptivity}.

\subsubsection{Cross-Paradigm Synthesis}
Table~\ref{tab:daigaudit} synthesizes this paradigm audit. Across all three areas of research, existing technologies frame health management either as an isolated cognitive activity, a static pairwise link, or a dyadic human-algorithm exchange. \textit{Sangyadhi} demonstrates that collective care is defined precisely by the dynamic, interdependent negotiation of context, responsibility, trust-mediated visibility, and collaborative actionability.

\begin{table}[ht!]
\centering
\small
\begin{tabular}{@{}p{0.18\textwidth}p{0.20\textwidth}p{0.28\textwidth}p{0.26\textwidth}@{}}
\toprule
\textbf{Paradigm \newline \& Archetype} & \textbf{Dominant \newline Analytical Focus} & \textbf{Structural Omission \newline Revealed by Sangyadhi} & \textbf{Methodological \& \newline Analytical Implication} \\
\midrule

\textbf{Individual PI} \newline (e.g., Apple HealthKit)~\cite{10.1145/1753326.1753409, Apple_HealthKit} &
Individual self-monitoring, stages of reflection, personal behavior change goals &
Conflates sensing, interpretation, and action; erases domestic routines and treats sharing as passive broadcast &
Shift unit of analysis from individual dashboards to distributed multi-actor trajectories; audit invisible coordination labor \\

\midrule

\textbf{Relational/ Family} \newline (e.g., Health Connect, Flo, Clue)~\cite{10.1145/2998181.2998362, Google_HealthConnect, Flo_Partners} &
Interpersonal support, pairwise sharing, social modeling, data interoperability &
Assumes static permissions; fails to capture time-bounded trust, dynamic handoffs, and household power hierarchies &
Investigate ephemeral, purpose-bound visibility; differentiate task performance from legitimate decision-making authority \\

\midrule

\textbf{Clinical AI/ HCAI} \newline (e.g., CDSS, Predictive Health)~\cite{10.1145/3290605.3300233, 10.1145/3544548.3581251} &
Model accuracy, dyadic human-in-the-loop oversight, clinical workflow fit &
Dyadic blind spot: lacks role-aware alert routing, network-level explanations, contextual deferral, and contestability &
Model actionability as network-distributed enactment; evaluate which human is positioned, authorized, and supported to intervene \\

\bottomrule
\end{tabular}
\caption{\label{tab:daigaudit} Diagnostic audit across dominant digital health paradigms and platforms using the Sangyadhi framework.}
\end{table}

\subsection{Illustrative Diagnostic Walkthrough: Characterizing an Observed Care Context}
\label{illustration}

Having established how \textit{Sangyadhi} diagnoses macro-level research paradigms, we now demonstrate how field researchers can apply its constructs qualitatively to investigate an observed caregiving setting. Rather than presenting a software specification or a speculative code architecture, this walkthrough shows how the framework provides the descriptive scaffolding needed to determine whether an observed setting functions as a collective care context and to diagnostically identify its sociotechnical failure modes.

\subsubsection{The Field Context: Postnatal Care in an Urban Extended Family}
Consider an ethnographer or qualitative health researcher observing a multi-generational household in urban India during the postpartum period, a setting directly informed by documented sociotechnical dynamics in prior healthcare fieldwork~\cite{10.1145/3274344, 10.1145/3544548.3581553, srinivasan2025unpacking}. The care ecology comprises three primary human actors and an emerging suite of digital health tools:
\begin{itemize}
\item \textit{The Mother (Primary Care Recipient):} Recovering from childbirth, managing infant breastfeeding, and tracking postpartum recovery.
\item \textit{The Mother-in-Law (Domestic Caregiver):} Managing household operations, preparing traditional postpartum nutrition, assisting with infant sleep and bathing, and exerting substantial cultural authority over care rituals.
\item \textit{The Visiting Midwife / Community Health Worker (Clinical Intermediary):} Conducting periodic home visits, monitoring infant weight gain and neonatal jaundice, and serving as the liaison to formal clinical infrastructure.
\end{itemize}

While presented here as an illustrative walkthrough to demonstrate the framework's runtime mechanisms, this specific care ecology\textemdash characterized by the generational authority of the mother-in-law, the intermediary clinical role of the community health worker, and the lived temporalities of postnatal rituals\textemdash is synthesized directly from extensively documented empirical realities in CSCW and HCI4D literature~\cite{10.1145/3274344, 10.1145/3359272, 10.1145/3544548.3581553, sehgal2026we}.

\subsubsection{Diagnostic Evaluation: Is This a Collective Care Context?}
A superficial inquiry might categorize this household simply as an individual mother receiving informal domestic help. Applying the \textit{Sangyadhi} diagnostic criteria reveals that this environment is structurally a \textbf{collective care context}, exhibiting four interdependent properties:
\begin{enumerate}
\item \textit{Decoupled Sensing, Reflection, and Action:} The mother logs feeding times, but the mother-in-law observes physical hydration cues (such as skin turgor and fontanelle firmness), while the midwife tracks physiological weight curves. Sensing, interpretation, and action are structurally decoupled across an ensemble with distinct literacies and authority~\cite{10.1145/353485.353487}.
\item \textit{Temporal and Ritual Mediation:} Daily care practices are bound to domestic temporalities, including non-negotiable afternoon rest periods (\textit{postnatal rest}), dietary schedules, and infant massage rituals. System interventions cannot occur at arbitrary times without disrupting domestic equilibrium~\cite{38407}.
\item \textit{Relational Asymmetry and Bounded Visibility:} The mother shares routine infant logs openly with the mother-in-law, but maintains strict privacy boundaries regarding her own physical healing and emotional state. Meanwhile, clinical data is shared with the midwife only during specific evaluative windows, requiring ephemeral, purpose-bound visibility rather than permanent exposure of data~\cite {Nissenbaum_2004, 10.1145/3706598.3714089}.
\item \textit{Distributed Enactment and Authority:} If an anomaly is identified (e.g., infant dehydration), an automated alert delivered solely to the mother's phone is ineffective if she lacks the cultural autonomy or transport resources to seek clinical care independently. Action requires negotiated escalation across the mother, mother-in-law, and midwife~\cite{10.1145/3274344, 10.1145/3544548.3581553}.
\end{enumerate}

\subsubsection{Sociotechnical Care-Network Mapping Profile}
To critically document such environments without reducing human care to brittle code, researchers can construct a \textit{Sociotechnical Care-Network Mapping Profile}. Table~\ref{tab:caremap} illustrates how the four constructs structure qualitative field observations into a coherent descriptive model.

\begin{table}[ht!]
\centering
\small
\begin{tabular}{@{}p{0.22\textwidth} p{0.74\textwidth}@{}}
\toprule
\textbf{Construct Layer} & \textbf{Observed Sociotechnical Dynamics and Field Manifestations} \\
\midrule

\textbf{Foundational Layer:} \newline Contextual Grounding &
\textbf{Temporal Rhythms:} Low-contact domestic rest window observed between 14:00--16:00; routine checks aligned with morning bathing (08:00--10:00). Non-urgent digital interruptions during rest periods induce stress and abandonment. \newline
\textbf{Cultural Rituals:} Dietary practices and infant massage routines dictate care priority over clinical mobile app checklists. \\

\midrule

\textbf{Foundational Layer:} \newline Distributed Responsibility &
\textbf{Role Differentiation:} Mother holds experiential authority over infant suckling; Mother-in-law holds domestic operational authority over daily routines; Midwife holds clinical escalation authority. \newline
\textbf{Handoff Triggers:} Responsibility for infant soothing transfers to the mother-in-law when maternal fatigue scores exceed operational capacity; clinical responsibility transfers to the midwife upon physiological anomaly. \\

\midrule

\textbf{Mediating Layer:} \newline Trust-mediated Sharing &
\textbf{Granularity Boundaries:} Mother grants full access to infant feeding logs to the household, but restricts maternal perineal healing and postpartum mood data to personal views. \newline
\textbf{Ephemeral Clinical Visibility:} Midwife is granted 24-hour temporary access to high-resolution weight and hydration data upon an anomaly flag, automatically reverting to summary views upon resolution. \\

\midrule

\textbf{Enactment Layer:} \newline Collaborative Actionability &
\textbf{Role-Aware Routing:} Algorithmic alerts regarding dehydration risk ($Score \ge 0.7$) are routed simultaneously to the midwife (with clinical diagnostic traces) and the mother (with plain-language hydration steps). \newline
\textbf{Contextual Deferral \& Escalation:} Non-critical alerts are held during afternoon rest windows; unacknowledged urgent alerts escalate from SMS to an interactive voice response (IVR) call to the secondary caregiver. \\

\bottomrule
\end{tabular}
\caption{\label{tab:caremap} Sociotechnical care-network mapping profile for the postnatal collective care context.}
\end{table}

\subsubsection{Analyzing Trajectories and Failure Modes: An Autopsy of Breakdown}
Applying this profile allows researchers to conduct an analytical autopsy of typical technology failures in multi-actor settings. Consider a scenario in which a mobile health app detects a 7.5\% weight loss in an infant over three days.

In a conventional, individual-centered system (Paradigm 1), an automated notification (``\textit{Warning: Infant weight loss exceeds threshold. Consult clinic immediately.}'') is pushed to the mother's personal device at 14:30. Through the \textit{Sangyadhi} lens, we can pinpoint precisely why this intervention fails:
\begin{enumerate}
\item \textit{Contextual Violation:} It interrupts the mandatory postpartum rest window (14:00--16:00), triggering acute anxiety and maternal guilt.
\item \textit{Responsibility Mismatch:} It treats the isolated mother as the sole actor, ignoring that she cannot leave the home independently and that the mother-in-law oversees household transportation and resources.
\item \textit{Actionability Failure:} The alert offers no clinical explanation or escalation path, leaving the family to debate the validity of the digital scale versus traditional somatic assessments of infant wellness.
\end{enumerate}

Under a \textit{Sangyadhi}-informed analysis, the researcher identifies the necessary sociotechnical coordination: the system recognizes the rest window, defers non-emergency communication, and upon threshold confirmation, issues a synchronized, role-aware handoff. It grants the midwife temporary, purpose-bound visibility into the 3-day weight history, alerts the midwife with clinical provenance, and routes a calm, actionable prompt to the household, suggesting an upcoming home visit. By analyzing the breakdown across all four lenses, researchers can diagnose whether failures stem from contextual friction, misallocated responsibility, privacy violations, or actionability bottlenecks.

\subsection{Methodological Blueprint: A Field Inquiry Protocol for Collective Care}
\label{checklist}

Building upon the paradigm audit and diagnostic walkthrough, we synthesize \textit{Sangyadhi} into an empirical field inquiry protocol for HCI researchers. This blueprint guides researchers through two critical phases of investigation: (1) establishing whether an observed setting is a collective care context, and (2) conducting qualitative fieldwork to investigate its dynamics.

\subsubsection{The Diagnostic Threshold: Differentiating Collective Care from Interpersonal Informatics}
A persistent methodological challenge in CSCW and health informatics is the conflation of \textit{interpersonal presence} with \textit{collective care}. Having multiple family members view an app or offer verbal encouragement does not necessarily constitute a collective care ecology. To establish analytical rigor, we propose that a field setting constitutes a \textbf{Collective Care Context} if and only if it satisfies three structural criteria:
\begin{itemize}
\item \textbf{Decoupled Agency:} Health information work (sensing, logging, interpreting, deciding, enacting) is structurally distributed across two or more actors, such that no single individual performs the entire care trajectory alone.
\item \textbf{Dynamic Responsibility Trajectories:} Operational duties and legitimate decision-making authority are subject to negotiation, handoffs, or escalation based on changing acute needs or caregiver availability.
\item \textbf{Interdependent Information Boundaries:} Information access cannot be satisfied by static, universal permissions; data visibility directly impacts relational trust, vulnerability, and the substantive capability of actors to provide care.
\end{itemize}

\subsubsection{The Qualitative Field Inquiry Matrix}
When investigating environments that meet this threshold, researchers can utilize the \textit{Sangyadhi Field Inquiry Matrix} (Table~\ref{tab:fieldinq}) to formulate interview guides, observational protocols, and participatory design probes.

\begin{table}[ht!]
\centering
\small
\begin{tabular}{@{}p{0.18\textwidth} p{0.22\textwidth} p{0.28\textwidth} p{0.26\textwidth}@{}}
\toprule
\textbf{Construct Lens} & \textbf{Core Diagnostic \newline Question} & \textbf{Observational Probes \newline \& Inquiry Prompts} & \textbf{Analytical Failure Modes if Overlooked} \\
\midrule

\textbf{Contextual \newline Grounding} &
How do local temporalities, routines, and cultural rituals shape the appropriateness of care? &
Observe domestic routines, quiet hours, and religious practices. Probe: \textit{``When during the day is health talk welcomed, and when is it intrusive? How do family rituals take precedence over medical tasks?''} &
Treating health management as an abstract, continuous duty; designing tools that disrupt domestic equilibrium and cause abandonment. \\

\midrule

\textbf{Distributed \newline Responsibility} &
How are cognitive labor, operational duty, and decision authority distributed and renegotiated? &
Map the division of labor across the family and intermediaries. Probe: \textit{``Who reads the numbers versus who decides what to do? What happens when the main caregiver falls ill or leaves the home?''} &
Conflating app access with legitimate authority; overburdening secondary actors with invisible coordination and emotional labor. \\

\midrule

\textbf{Trust-mediated \newline Sharing} &
How do relational boundaries, stigma, and vulnerability govern information visibility? &
Trace boundaries of modesty and stigma. Probe: \textit{``What health details are deliberately concealed from specific relatives? Under what emergency circumstances does private data become shareable?''} &
Implementing static, binary sharing toggles; enabling domestic surveillance, privacy leaks, and relational friction. \\

\midrule

\textbf{Collaborative \newline Actionability} &
How do automated insights and external guidance translate into coordinated network action? &
Observe how clinical advice circulates through household hierarchies. Probe: \textit{``When the app issues an alert, who sees it first, who decides if it is trustworthy, and who carries out the remedy?''} &
The dyadic blind spot; sending alarms to actors who lack power or literacy to act; creating alarm fatigue and unmanaged anxiety. \\

\bottomrule
\end{tabular}
\caption{\label{tab:fieldinq} The Sangyadhi qualitative field inquiry matrix for diagnosing collective care contexts.}
\end{table}

\subsubsection{Diagnostic Auditing of System Deployments}
Finally, when evaluating interactive technologies deployed in collective care settings, researchers can apply five diagnostic auditing heuristics:
\begin{enumerate}
\item \textbf{Map the Care Network:} Identify all human and nonhuman actors involved in health work, explicitly distinguishing technical access from legitimate social authority.
\item \textbf{Audit Temporal and Contextual Fit:} Evaluate whether interaction timings, notifications, and task demands align with local domestic rhythms and cultural practices or impose disruptive external schedules.
\item \textbf{Trace Responsibility and Handoff Paths:} Assess whether the system maintains legibility of who is currently accountable, provides lightweight handoff mechanisms, and avoids shifting invisible coordination labor onto marginalized actors.
\item \textbf{Evaluate Trust and Visibility Lifecycles:} Examine whether data sharing supports time-bounded, purpose-specific, and ephemeral access with clear provenance and revocation mechanisms.
\item \textbf{Scrutinize Multi-Actor Actionability:} Trace how recommendations and automated inferences are routed, explained, deferred, or contested, ensuring that interventions reach actors who possess both the authority and the capacity to act.
\end{enumerate}

By structuring empirical inquiry around these dimensions, \textit{Sangyadhi} provides the HCI community with a critical methodology for understanding, evaluating, and conceptualizing technologies that respect the collective reality of care.
\section{Discussion, Implications, and Challenges}
\label{discussion}

\textit{Sangyadhi} makes a bounded theoretical and methodological claim: investigating and designing for collective care requires the HCI community to treat context, responsibility, information sharing, and coordinated action as interdependent sociotechnical concerns. The framework does not claim to prescribe a universal configuration of care, nor does it suggest that technical mechanisms alone can resolve entrenched health inequities. Instead, it provides a principled diagnostic vocabulary for characterizing multi-actor care ecologies, identifying structural omissions in existing systems, and structuring qualitative field inquiry.

\subsection{Theoretical Contribution}
\label{claims}

\textit{Sangyadhi}'s primary theoretical contribution is the synthesis of established perspectives that are typically applied to isolated dimensions of human-technology interaction. Situated agency foregrounds how action is constituted through social, temporal, and material circumstances~\cite{38407, emirbayer1998agency}; distributed cognition explains how cognitive activity is coordinated across people and artifacts~\cite{10.1145/353485.353487}; the capabilities approach emphasizes substantive opportunities for meaningful participation and decision-making~\cite{sen1999development, Nussbaum_2011}; and actor-network theory conceptualizes agency as emerging through heterogeneous human and nonhuman configurations~\cite{Callon_1984, Latour_2005}.

\textit{Sangyadhi} translates these foundational perspectives into four interdependent analytical constructs: \textit{Contextual Grounding}, \textit{Distributed Responsibility}, \textit{Trust-mediated Sharing}, and \textit{Collaborative Actionability}. The contribution lies squarely in their integration. While prior HCI scholarship provides rich accounts of distributed activity, contextual privacy, values, infrastructuring, and human-AI oversight in isolation~\cite{engeström2015learning, Star_Ruhleder_1996, Pipek_Wulf_2009, Nissenbaum_2004, HCI-015, 10.1145/3290605.3300233, 10.1145/3411764.3445088}, these concerns have remained analytically fragmented. \textit{Sangyadhi} articulates their structural interdependence: situated context defines legitimate agency; division of authority determines who requires information visibility; trust-mediated visibility enables or constrains participation; and the resulting network configuration dictates whether an intervention can be collaboratively enacted. This integrated lens distinguishes \textit{Sangyadhi} from frameworks that analyze privacy, explainability, or interoperability as separate technical features.

\subsection{Practical and Methodological Implications}
\label{practical}

For researchers and practitioners, \textit{Sangyadhi} shifts the focus of health informatics from individual data pipelines toward the social organization of care. Existing data interoperability approaches (such as FHIR standards and unified data models) successfully make health metrics portable across systems, but they remain agnostic to interpersonal coordination and social governance~\cite{10.1145/3402524, 10.1145/3706598.3714069}. \textit{Sangyadhi} provides the analytical criteria needed to bridge this gap.

Methodologically, the framework equips field researchers with a structured protocol for empirical investigation. As outlined in Section~\ref{checklist}, the \textit{Diagnostic Threshold} prevents researchers from conflating simple interpersonal presence with collective care, while the \textit{Field Inquiry Matrix} (Table~\ref{tab:fieldinq}) structures observational protocols, contextual interviews, and participatory design probes. By guiding researchers to examine temporal boundaries, authority handoffs, and communicative hierarchies, \textit{Sangyadhi} complements established traditions in Participatory Action Research (PAR) and Ethnographic Action Research (EAR)~\cite{PEREIRA201566, Baum854, 10.1108/IJMPB-09-2021-0232, BetterEvaluation_EARToolbox_2017, black2021ethnographic, morgan2016ethnographic}.

For system evaluation and design, \textit{Sangyadhi} broadens the criteria by which health technologies are assessed. Algorithmic accuracy, processing latency, and interface usability remain essential, but they cannot determine whether a system respects household quiet windows, clarifies who is accountable for an abnormal trend, or allows caregivers to contest automated advice. Research on automation bias, invisible labor, and clinical coordination demonstrates that technical precision alone frequently introduces severe sociotechnical friction~\cite{10.1109/ICHI.2015.26, 10.1145/3491102.3502104, 10.1145/3610218, ulloa2022invisible, 10.1145/3643834.3660707, kucking2024automation}. Applying \textit{Sangyadhi}'s diagnostic auditing heuristics enables designers to anticipate and mitigate these breakdowns before longitudinal deployment.

\subsection{Ethical Safeguards and Boundary Management}
\label{ethics}

Making informal roles, family responsibilities, and information flows legible within digital systems can facilitate coordination, but it also carries inherent risks of domestic surveillance, coercion, and deepened marginalization. Privacy scholarship, value-sensitive design, feminist HCI, and standpoint theory have long warned that technical systems can formalize and reinforce existing power asymmetries~\cite{Nissenbaum_2004, HCI-015, 10.1145/3706598.3713950, Hartsock1983}. \textit{Sangyadhi} therefore treats ethical boundary management not as an optional configuration, but as a core requirement of collective care inquiry and design:

\begin{itemize}
\item \textit{Least Privilege and Ephemeral Visibility:} Information sharing must be purpose-bound, temporary, and revocable. Role-based models, dynamic consent interfaces, and bounded disclosure mechanisms provide essential foundations for preventing relational surveillance~\cite{10.1145/3613904.3642728, 10.1145/3613904.3642815, 10.1145/3613904.3642851, 10.1145/3706598.3714089, 10.1145/3706598.3714019, 10.1145/2441776.2441838}.
\item \textit{Contestability and Meaningful Oversight:} Consequential interventions must remain contestable. Caregivers and patients must have the substantive capacity to challenge automated alerts, defer non-critical prompts, and override algorithmic suggestions~\cite{10.1145/3290605.3300233, shneiderman2022human, 10.1145/3706598.3713184}. To avoid the ``formalization paradox''\textemdash where rigid digital tracking over-bureaucratizes fluid domestic care\textemdash systems must accommodate ``graceful ambiguity,'' enabling informal care overrides without logging administrative failures.
\item \textit{Anti-Surveillance Provenance:} Provenance tracking must serve mutual accountability and comprehension rather than domestic policing. Participants should understand why data was shared or altered without turning audit logs into instruments of familial scrutiny~\cite{10.1145/3706598.3714019, 10.1145/2675133.2675279}.
\item \textit{Equity and Accessibility:} Inquiry and design must account for structural disparities in literacy, gendered autonomy, language, and technology access. Plain-language explanations, audio-first interfaces, and mediated consent are vital to ensuring that governance mechanisms do not exclude peripheral or vulnerable caregivers~\cite{chronaki2017search, O_Connor2025, xu2025impact, october2023_gender_digital_health, s2025whotimeunderstandingreceptivity, choi2025bridging}.
\item \textit{Revisability of Contextual Norms:} Contextual rules must remain dynamic and renegotiable. Treating domestic routines or cultural practices as immutable parameters risks reinforcing oppressive household hierarchies. Ongoing participatory engagement is necessary to ensure encoded norms adapt as family relationships evolve~\cite{Star_Ruhleder_1996, Pipek_Wulf_2009, PEREIRA201566, 10.1145/3706598.3713436, 10.1145/3613904.3642727, 10.1145/3706598.3713755}.
\end{itemize}

\subsection{Methodological Challenges and Boundaries}
\label{challenges}

Applying an analytical framework to human care relations involves navigating several epistemological and methodological tensions:

First, caregiving relationships are inherently fluid, affective, and ambiguous. Actors frequently inhabit multiple overlapping roles, authority is constantly negotiated through unstated gestures, and domestic responsibilities resist rigid categorization. While \textit{Sangyadhi} provides structured constructs to make these dynamics legible, researchers and designers must caution against a false sense of precision. As distributed cognition, activity theory, and infrastructuring scholarship emphasize, human coordination cannot be reduced to a static diagram~\cite{10.1145/353485.353487, engeström2015learning, Star_Ruhleder_1996, Pipek_Wulf_2009}. \textit{Sangyadhi}'s constructs should be used as sensitizing concepts for qualitative inquiry rather than as rigid boundaries.

Second, contextual grounding carries an inherent risk of \textit{cultural ossification}. Categorizing domestic routines or traditional practices can inadvertently reify conservative gender roles or privilege dominant family members over vulnerable care recipients~\cite{10.1145/3706598.3713950, Hartsock1983, O_Connor2025}. Field researchers utilizing this framework must maintain reflexivity, actively probing who benefits and who is burdened by existing cultural arrangements.

Third, collective care ecologies vary dramatically across socioeconomic and institutional settings. While this paper illustrated multi-actor care through a multi-generational postpartum household, other settings (such as transient migrant networks, stigma-laden chronic illness support, or resource-constrained public clinics) may exhibit vastly different patterns of trust, literacy, and mobility. Community health technologies and participatory methodologies provide critical guidance for tailoring the framework's lenses to these marginalized ecologies~\cite{10.1145/3274344, 10.1145/3274345, 10.1145/3313831.3376170, Baum854, 10.1108/IJMPB-09-2021-0232}.

Finally, as an analytical framework, \textit{Sangyadhi} is designed to guide inquiry, critique, and conceptualization; it does not replace the necessity of sustained, participatory engagement with communities. While the problem dimensions synthesized here are grounded in extensive empirical findings across digital health and proactive collective care~\cite{srinivasan2025unpacking}, ongoing empirical work is needed to examine how researchers utilize these diagnostic tools in the field, where their analytical lenses require refinement, and how they can best support the conceptualization of equitable care technologies.
\section{Conclusion}
\label{conclusion}

For over two decades, digital health research in HCI has been dominated by a quietly pervasive paradigm: the premise that health management is fundamentally an individual enterprise. PI tools continue to treat users as isolated agents operating individual tracking devices that report to personal dashboards. While the emergence of Family Informatics and relational sharing features has successfully drawn attention to social presence, the field has continued to lack a critical analytical grammar for determining what fundamentally constitutes a collective care context and for evaluating its sociotechnical breakdowns.

This paper introduced \textit{Sangyadhi}, an analytical sociotechnical framework for diagnosing, investigating, and conceptualizing collective care ecologies. Synthesizing situated action, distributed cognition, the capabilities approach, and actor-network theory, \textit{Sangyadhi} reframes the primary unit of analysis from the solitary user to the interdependent social arrangements that make care possible. 

The framework addresses our core research questions (RQ1-RQ4) through four interlocking analytical lenses. Rather than operating as isolated features, these lenses demonstrate that situating health practices within domestic rhythms (\textit{Contextual Grounding}) dictates how task performance and authority are disentangled (\textit{Distributed Responsibility}). This, in turn, governs how data visibility must be calibrated to relational vulnerability (\textit{Trust-mediated Sharing}), ultimately ensuring that automated insights can be appropriately routed, deferred, and contested across the care network (\textit{Collaborative Actionability}).

Through our diagnostic audit of a decade of HCI literature across individual personal informatics, relational sharing, and clinical AI, we demonstrated that existing technologies consistently falter not from a lack of data or algorithmic accuracy, but from structural omissions in how they represent shifting authority, domestic temporality, and collective action. By providing an explicit diagnostic threshold, an illustrative care-network mapping profile, and an operational field inquiry matrix, \textit{Sangyadhi} equips qualitative researchers and designers with the methodological scaffolding required to evaluate whether an observed setting operates as a collective care context, structure empirical fieldwork, and anticipate sociotechnical failure modes before technology is introduced.

Ultimately, \textit{Sangyadhi} establishes that supporting collective care is not a matter of adding secondary viewer dashboards or pairwise sharing toggles to systems built for individuals. It requires recognizing that human care is an ongoing, relational accomplishment enacted through fragile ecologies of people, routines, infrastructures, and moral obligations. By providing a principled vocabulary to render these dynamics legible and negotiable, \textit{Sangyadhi} offers an analytical foundation for HCI to look beyond the quantified self and take the collective nature of care seriously.

\bibliographystyle{ACM-Reference-Format}
\bibliography{references}
\clearpage %%% Uncomment this line and comment out the ``thebibliography'' section below to use the external .bib file (using bibtex) .





% \end{thebibliography}

\end{document}